\documentclass{article}
\usepackage{subcaption}
\usepackage{graphicx}
\usepackage{authblk}
\newcommand{\bfr}{\begin{flushright}}
\newcommand{\efr}{\end{flushright}}
\usepackage{xspace}
\newcommand{\pstar}{p^{(\ast)}}
\newcommand{\calchep}{\texttt{CalcHEP}\xspace}
\newcommand{\exhume}{\texttt{ExHuMe}\xspace}
\newcommand{\lpair}{\texttt{LPAIR}\xspace}
 
\begin{document}
\title{Exclusive processes in proton-proton collisions with the CMS experiment at the LHC
\date{}
\thanks{Presented at  the Low x workshop, May 30 - June 4 2013, Rehovot and
Eilat, Israel}%
}
\author{Laurent Forthomme\footnote{\texttt{laurent.forthomme@uclouvain.be}}, on behalf of
  the CMS collaboration
\\
{\small
  Centre for Cosmology, Particle Physics and Phenomenology,
}\\
{\small
  Universit\'e catholique de Louvain,
}\\
{\small
  Chemin du Cyclotron 2, 1348 Louvain-la-Neuve, Belgium
}
\smallskip\\
}
\date{\today
}
\maketitle
\begin{abstract}
We present the recent measurements of exclusive processes performed in the CMS
experiment at the LHC using data collected at a centre of mass energy of 7 TeV.
These measurements include the double-pomeron production of photon pairs, the
two-photon production of leptons pairs, and the previously undetected two-photon
production of $W$ boson pairs.
While in case of the two first processes that enables to set limits on
production cross-section, in the later case it provides also stringent limits on
the anomalous quartic gauge couplings.
\\
\end{abstract}

\section{Introduction}
Central exclusive processes provide an interesting field of study in particle
physics. With its excellent performance, the CMS experiment, whose full
description can be found elsewhere\cite{JINST,PTDR2}, has managed by now to make
a number of significant observations of these processes, hence to probe the
Standard model in a unique way.
A striking signature is expected for such processes, with two very forward
(though undetected) scattered protons and two large rapidity gaps between these
protons and the central exclusively produced system. An overview is hereby
presented for some of these results, such as the search for central exclusive
production of photon pairs, the exclusive two-photon production of lepton pairs,
and the achieved limits on anomalous quartic gauge couplings in the two-photon
production of $W$ pairs.

\begin{figure}[h!]
  \centering
  \begin{subfigure}[b]{.33\textwidth}
    \centering
    \includegraphics[width=\textwidth]{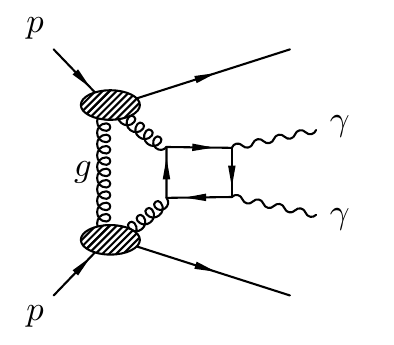}
    \caption{}
    \label{fig:gluglu}
  \end{subfigure}
  \begin{subfigure}[b]{.32\textwidth}
    \centering
    \includegraphics[width=\textwidth]{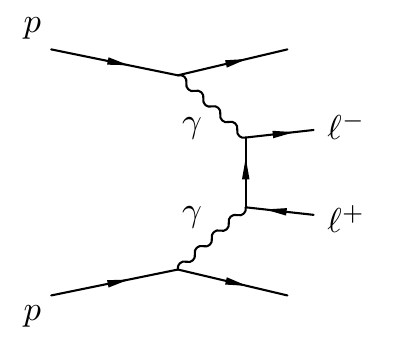}
    \caption{}
    \label{fig:gamgam}
  \end{subfigure}
  \begin{subfigure}[b]{.33\textwidth}
    \centering
    \includegraphics[width=\textwidth]{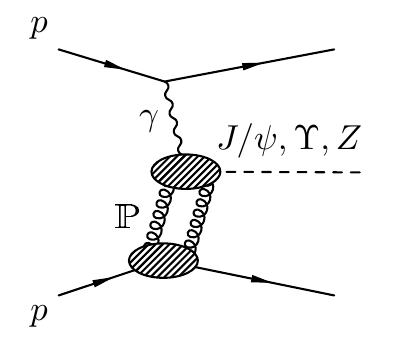}
    \caption{}
    \label{fig:photpom}
  \end{subfigure}
  \caption{\small Various types of exclusive processes involving (a) the double
    pomeron exchange, (b) the two-photon production, and (c) the photon-pomeron
    fusion.}
  \label{fig:gg-comparison-exptheor}
\end{figure}

\section{Central-exclusive $\gamma\gamma$ production
\label{sec2}}

The search for central exclusive production of two-photon events as represented
on Fig. \ref{fig:gluglu}, constitutes a direct test for the QCD predictions
involving the so-called double pomeron exchange.
In perturbative calculations this process can be interpreted as a
$gg\to\gamma\gamma$ subprocess involving an additional low-momentum gluon
exchange enabling the cancellation of the colour flow involved by the
interacting gluons.

With a data sample corresponding to an integrated luminosity of
$36~\mathrm{fb}^{-1}$ collected in 2010 by the CMS detector, the limit on the
cross-section of central exclusive production of two photons can be quoted for
the first time at a centre of mass energy $\sqrt s=7~\mathrm{TeV}$ after a
measurement performed at $\sqrt s=1.96~\mathrm{TeV}$ by the CDF experiment at
the Tevatron\cite{Aaltonen:2011hi}.
The candidate events are requested to contain two energetic photons with a
transverse energy higher than $5.5~\mathrm{GeV}$ as measured by the
electromagnetic calorimeter within its acceptance (corresponding to a
pseudo-rapidity range $|\eta|<2.5$). These two photons are furthermore expected
to be balanced in their transverse deposited energy and be back-to-back in
azimuthal angle.
The exclusivity condition is then satisfied when no additional activity (apart
from the two photons) is observed in the full detector (thus in the range
$|\eta|<5.2$). This allows a reduced observation of the proton-dissociative
events, when one or both forward scattered proton dissociate in a hadronic
system hence additional activity is expected in the forward region of the
detector.

\begin{figure}[h!]
  \centering
  \includegraphics[width=.504\textwidth]{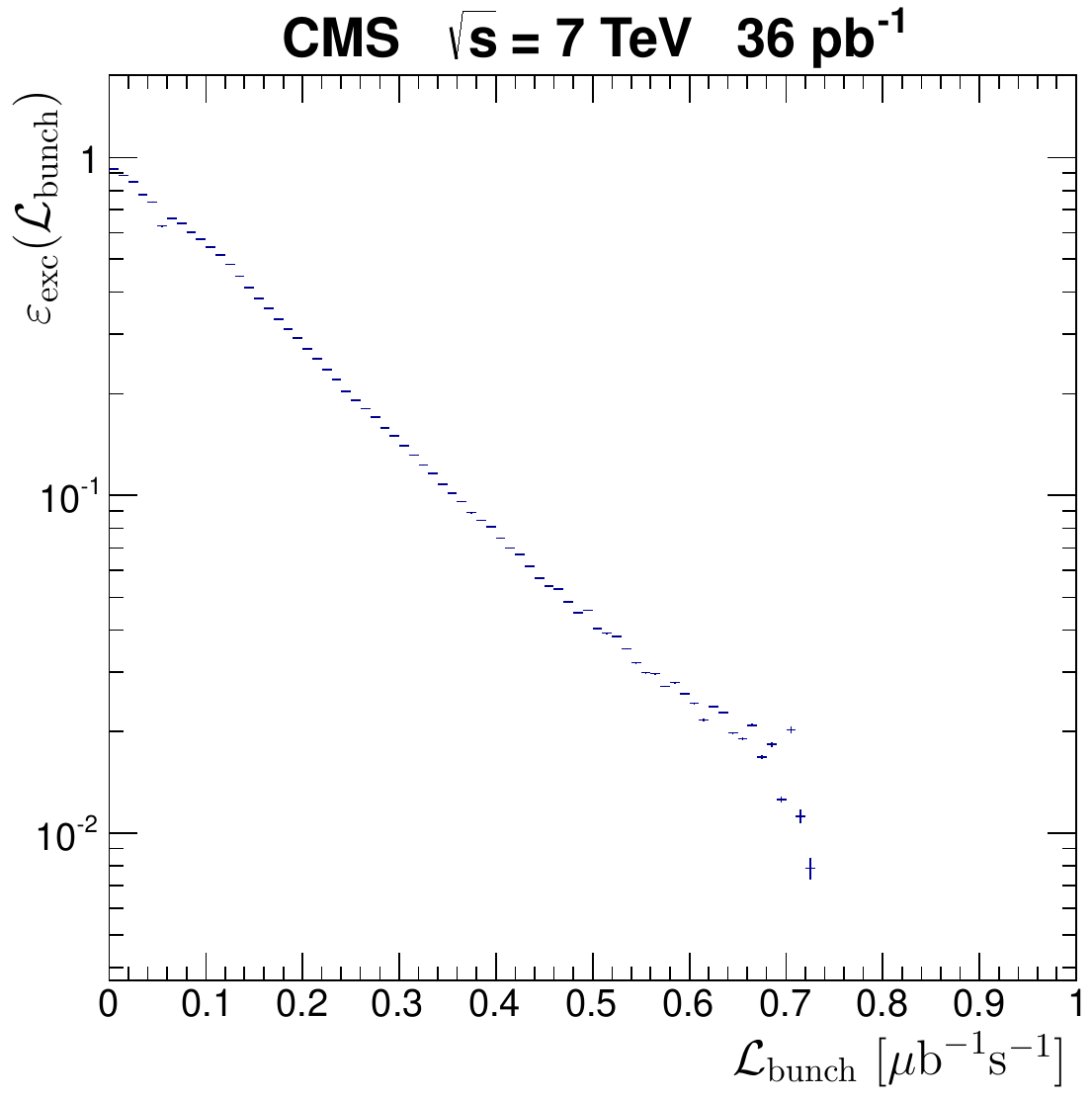}
  \caption{\small Exclusivity condition efficiency with respect to the
    instantaneous luminosity of each protons bunch. This exclusivity efficiency
    is computed to be the ratio between the number of events passing the
    selection criteria and the total number of the so-called \emph{zero bias}
    events collected in 2010 at $\sqrt s=7~\mathrm{TeV}$. Figure from
    \cite{Chatrchyan:2012tv}.}
  \label{fig:gg-exclueff}
\end{figure}

One of the sources of inefficiency considered in the analysis is due to the
relatively high instantaneous luminosity delivered by the LHC machine, whose
main drawback is the multiplication of the primary vertices in each event. This
phenomenon, the ``events pileup'', is expected to lower in a significant way the
efficiency of the exclusivity condition with respect to the luminosity.
This is shown in Fig. \ref{fig:gg-exclueff} in which the efficiency is evaluated
using \emph{zero bias} events collected parasitically during the data-taking.

No candidate events are observed with a background expectation of $1.79\pm 0.40$
events arising from the multiple processes involving an inclusive or exclusive
production of photons, such as pairs of misidentified photons (from exclusive
production of electron pairs) or from the exclusive $\pi^0\pi^0$ production. An
upper limit on the exclusive production cross-section can therefore be set
according to this result :
\begin{displaymath}
  \sigma(E_T(\gamma)>5.5~\mathrm{GeV}, |\eta(\gamma)|<2.5) < 1.18~\mathrm{pb}.
\end{displaymath}

This limit is then compared to the theoretical prediction from various
approaches using different orders in the perturbation theory, and parton
densities functions (PDFs).
A graphical view of this comparison can be found in Fig.
\ref{fig:gg-comparison-exptheor}, without the most recent prediction
\cite{HarlandLang:2012qz} computed at the leading order (LO) and the
next-to-leading order (NLO), providing a production cross section of
$0.180~\mathrm{pb}$ for the former using the \texttt{MSTW08}\cite{Martin:2009iq}
PDF, and $0.039~\mathrm{pb}$ for the latter using the \texttt{MRST99}
\cite{Martin:1999ww} PDF.
These signal predictions are computed using the \exhume 1.34\cite{Monk:2005ji}
event generator implementing the KMR model\cite{Khoze:2001xm} where the
proton-gluon couplings are determined perturbatively.
In this model, the two-photon system is then produced by the mean of a quark
box, as represented on Fig. \ref{fig:gluglu}.

\begin{figure}[h!]
  \centering
  \includegraphics[width=.51\textwidth]{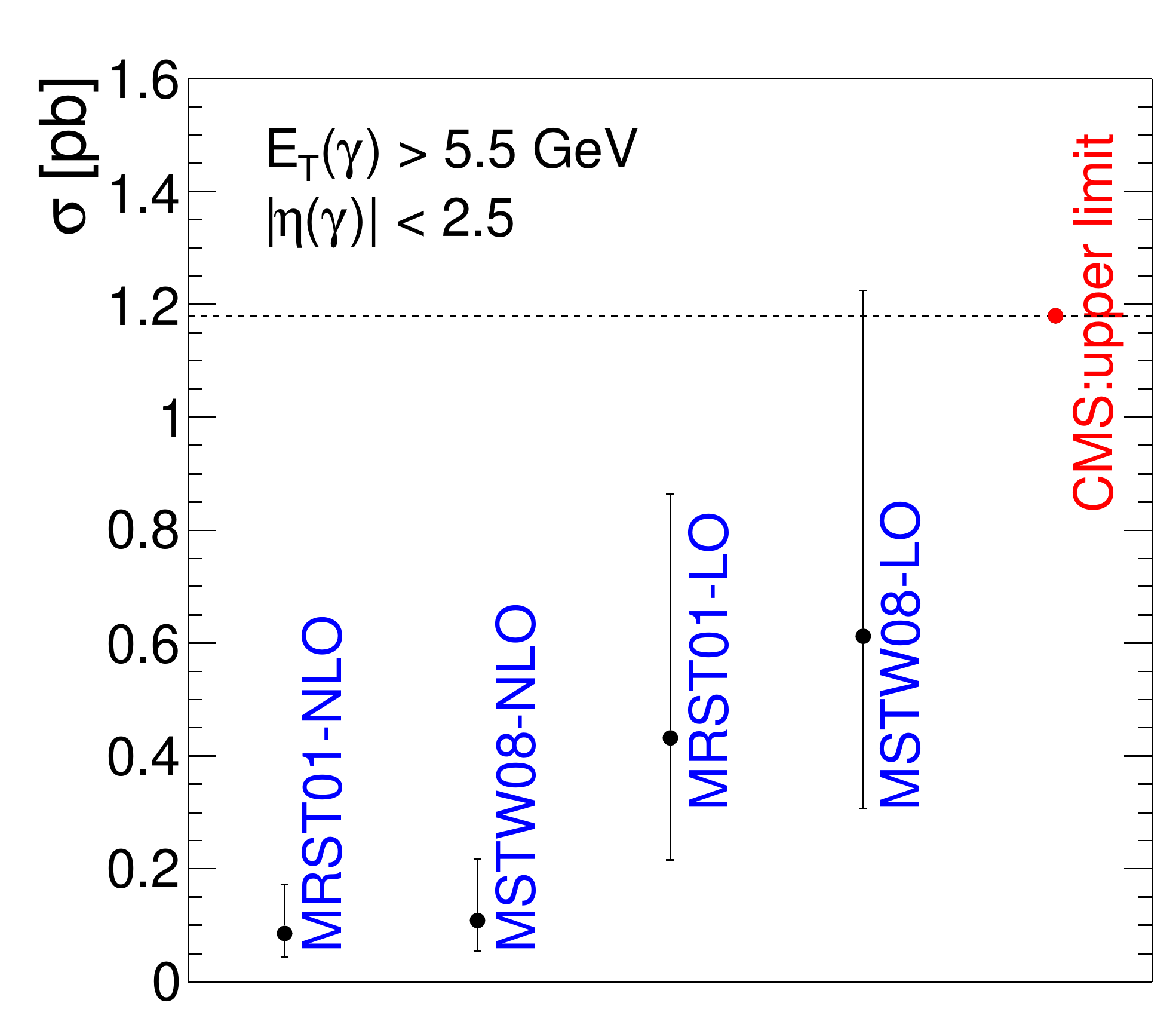}
  \caption{\small Comparison of the CMS experimental upper limit on the central
    diffractive production of photon pairs with the theoretical predictions at
    leading order (LO) and at the next to leading order (NLO) for different
    parton density functions.}
  \label{fig:gg-comparison-exptheor}
\end{figure}

This result shows that the theoretical predictions for this exclusive process,
mediated by gluons only, are in good agreement with the observed upper limit on
the cross-section.

\section{Exclusive two-photon production of lepton pairs
\label{sec3}}

Two different di-lepton exclusive analyses have been released using the data
collected in 2010 at $7~\mathrm{TeV}$, namely for the measurements of
$\gamma\gamma\to e^+e^-$\cite{Chatrchyan:2012tv} and of
$\gamma\gamma\to\mu^+\mu^-$\cite{Chatrchyan:2011ci}, such as represented on Fig.
\ref{fig:gamgam}.

The signal for such processes is simulated using the \lpair
\cite{Baranov:1991yq,Vermaseren:1982cz} Monte Carlo generator developed in the
1980s for the HERA $ep$ collider experiments. It uses the full matrix element
calculation for a $p p \to \pstar\ell^+\ell^-\pstar$ process, where the photon
coupling to a proton is described by the proton electromagnetic form-factors, in
case when a fully exclusive production is simulated and both incident protons
survived the interaction.
In the proton inelastic case the form-factor is replaced by the proton structure
functions from HERA fits. The rapidity gap survival probability (due to
re-scattering) is not modeled by \lpair, and it is set to unity in these
analyses.

The event selection is requiring two leptons which are energy- or
momentum-balanced and back-to-back in the transverse plane.
This corresponds to a $\left|p_T(\ell^+)-p_T(\ell^-)\right|<1~\mathrm{GeV}$ as
well as an acoplanarity describing the difference in azimuthal angles,
$\left|1-\Delta\phi(\ell^+,\ell^-)/\pi\right|<0.1$.
Furthermore, the dimuon analysis requires each of the two muons to carry a
transverse momentum larger than $4~\mathrm{GeV}$ in the range $|\eta(\mu)|<2.1$.
In order to reject the exclusive photoproduction of the low-mass resonances such
as the $J/\psi$, the $\psi'$, and the $\Upsilon(1s,2s,3s)$, an invariant mass
cut is applied $m(\mu^+\mu^-)>11.5~\mathrm{GeV}$.
For the dielectron analysis, a electron-positron pair with a transverse energy
deposit in the calorimeters $E_T>5.5~\mathrm{GeV}$ are selected in the range
$|\eta(e)|<2.5$. No additonal invariant mass cut is needed in this case.

\begin{figure}[h!]
  \centering
  \begin{subfigure}[b]{.46\textwidth}
    \centering
    \includegraphics[width=\textwidth]{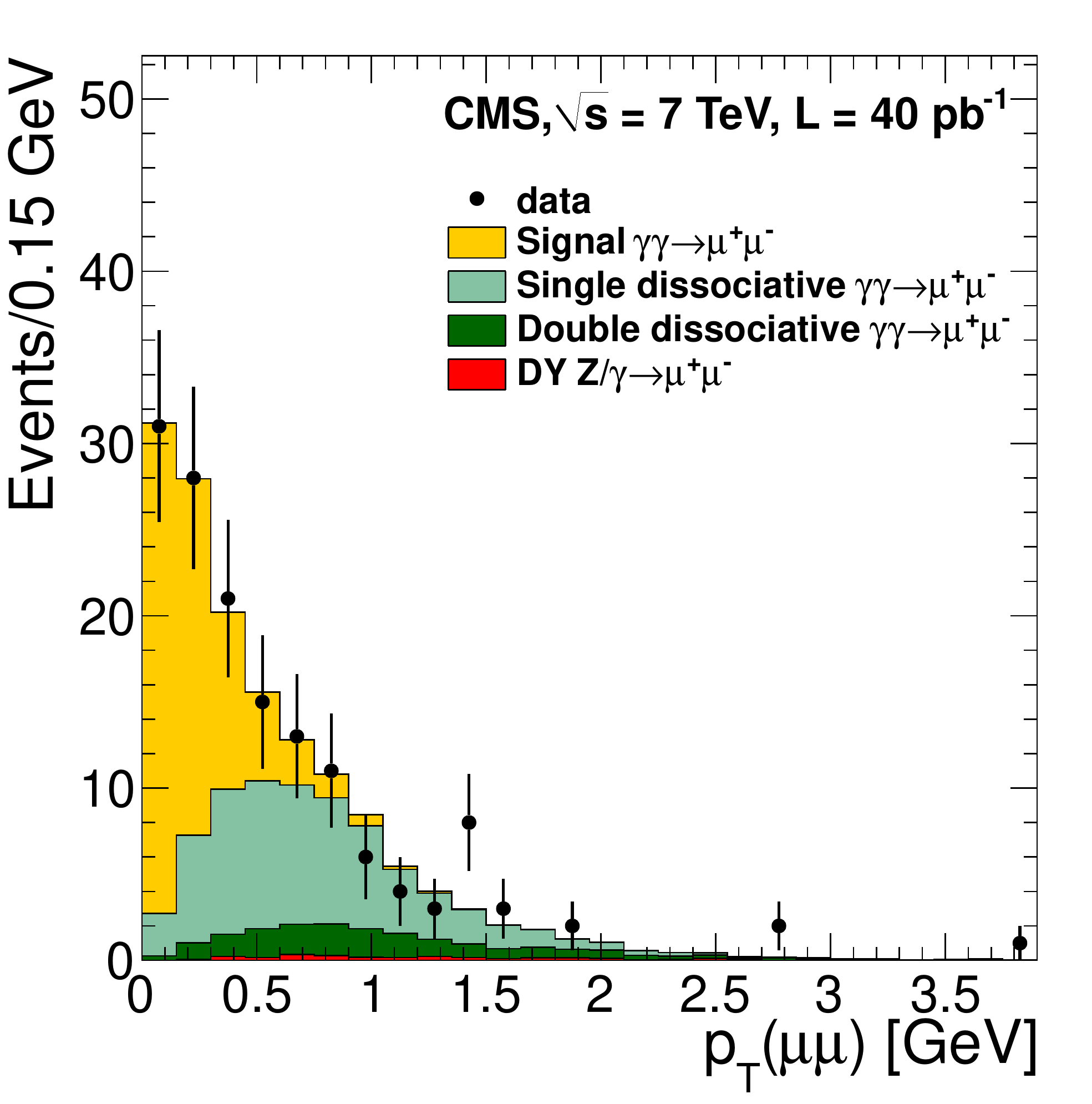}
    \caption{}
    \label{fig:ptpair-muon}
  \end{subfigure}
  \begin{subfigure}[b]{.49\textwidth}
    \centering
    \includegraphics[width=\textwidth]{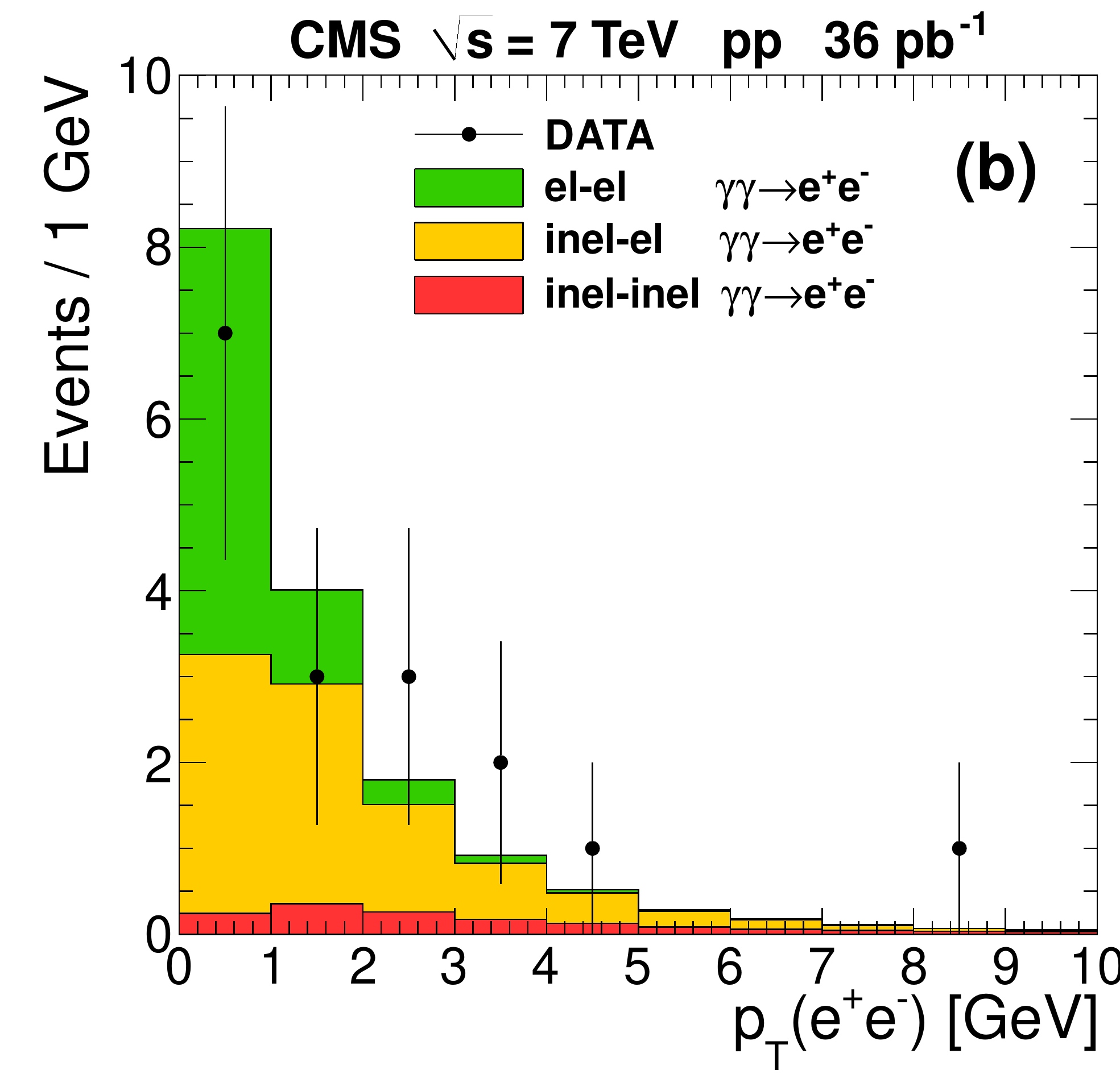}
    \caption{}
    \label{fig:ptpair-electron}
  \end{subfigure}
  \caption{\small Muons (left) and electron (right) pairs transverse momentum
    distributions for the candidates selected in the two-photon production of
    leptons pairs.}
  \label{fig:ptpair-ee-mm}
\end{figure}

To ensure the exclusivity condition, the dimuon analysis requires no additional
tracks within $2~\mathrm{mm}$ in the longitudinal direction around the dimuon
primary vertex, while the dielectron case rejects all events where any other
particles are reconstructed within the full acceptance of the detector,
$|\eta|<5.2$.
This very tight condition, along with a noticeable difference in the collected
integrated luminosity, results in large efficiency difference, hence in the
number of dimuon candidates (184 events) and the dielectron ones (17
candidates), as represented in Fig. \ref{fig:ptpair-ee-mm}.
Furthermore, the higher statistics provided by the dimuon channel allows an
extraction of the elastic signal contribution as well as the correction to be
applied on the theoretical yield predicted for the proton-dissociative part.
These corrections are determined from a binned maximum-likelihood fit to the
measured $p_T(\mu^+\mu^-)$ distribution as represented on Fig.
\ref{fig:ptpair-muon}.

These two results allow to improve the understanding of this purely
electromagnetic process, by observing 17 candidates for the dielectron channel
and by a measurement of a production cross-section at $\sqrt s=7$ TeV for the
dimuon channel :
\begin{displaymath}
  \sigma(pp\to p\mu^+\mu^-p) = 3.38{}^{+0.58}_{-0.55}~(\mathrm{stat.})\pm 0.16~(\mathrm{syst.})\pm 0.14~(\mathrm{lumi.})~\mathrm{pb}
\end{displaymath}
which is consistent with the theoretical prediction.

\newpage
\section{Two-photon production of $W$ pairs and limits on anomalous quartic gauge couplings
\label{sec4}}

Several processes beyond the Standard model predict an anomalous behaviour of
the quartic gauge coupling, such as new gauge bosons production or heavy quarks
exchanges\cite{Eboli:2003nq}.
These anomalous quartic gauge couplings (or AQGCs) can be translated into a
higher production rate, or discrepancies in the kinematic distributions of
multiple final state particles.

The LHC experiments have been predicted to be sensitive to such behaviours when
involving the two-photon interactions
\cite{deFavereaudeJeneret:2009db,Pierzchala:2008xc}.
Hence, a search for these anomalous couplings is performed at high energies in
the CMS experiment with the data collected in 2011 at $\sqrt s=7$ TeV, using the
challenging yet previously unobserved two-photon production of $W^\pm$ bosons
pairs\cite{Chatrchyan:2013foa}, where both the gauge bosons decay leptonically.

The channel of interest for this analysis is the different leptons flavours
decay, and especially $e^\pm\mu^\mp\nu\bar\nu$. Indeed, since the
high-statistics same-flavour channels are saturated with their main sources of
background, the $e^+e^-\nu\bar\nu$ and $\mu^+\mu^-\nu\bar\nu$ final states are
set aside in the current analysis.
For the former, the dominant sources of background are the inclusive
\emph{Drell-Yan} and the exclusive two-photon productions of $\tau$ leptons
pairs as well as the inclusive $W^+W^-$ production and the leptonic decay of
$t\bar t$ events.
On the other hand, the backgrounds for the later include the inclusive
\emph{Drell-Yan} production of same leptons pairs and exclusive
$\gamma\gamma\to\ell^+\ell^-$ (as seen in section \ref{sec3}), which are
predicted to be more than one order of magnitude larger than the exclusive
$\gamma\gamma\to W^+W^-$.

The two neutrinos being left undetected, the candidates are to contain two
leptons with $p_T(\ell)>20$ GeV reconstructed within the full detector
acceptance, such as $|\eta(\ell)|<2.4$, and matched to one single primary vertex
from which no additional tracks originates. Indeed, the high luminosity
conditions from the 2011 runs of the LHC giving rise to multiple interactions
within the same bunch crossing (the so-called "pileup"), this condition on the
additional tracks is the only one ensuring a sufficient efficiency in the
selection.

The leptons pair is required to have an invariant mass higher than $20$ GeV,
and two kinematic regions can be built to isolate the Standard model or the
anomalous quartic gauge couplings search regions.
For the former, a transverse momentum $p_T(e^\pm\mu^\mp)>30$ GeV is imposed on
the leptons pair, while the later is more stringent with a lower cut of $100$
GeV.

In order to simulate the theoretical prediction of the Standard model and the
anomalous scenarios in these two regions, extra diagrams have to be taken into
account in the final computation.
These additional processes are the protons single- and double-dissociative cases
which cannot be untangled from the purely elastic contribution without the
information on the outgoing protons.
Hence, an estimation of these two contributions is given by using the high
statistics $\mu^+\mu^-$ final state probed in the same phase space and
extracting a "scale factor" which can then be applied on the $W^+W^-$ signal :
\begin{displaymath}
  F = \left.\frac{n(\mu\mu~\mathrm{data})-n(\mu\mu~\mathrm{background})}{n(\mu\mu~\mathrm{elastic})}\right|_{m(\mu\mu)>160~\mathrm{GeV}} = 3.23\pm 0.53~\mathrm{(stat. + syst.)},
\end{displaymath}
with the background events defined as the prediction of inclusive
\emph{Drell-Yan} production of muons and taus pairs, and the elastic
$\gamma\gamma\to\mu^+\mu^-$ events number predicted by \lpair.

Given that rescaling in the probed high mass region a theoretical production
cross-section is extracted for two kinematic search regions using the \calchep
\cite{Pukhov:2004ca} generator. The theoretical prediction cross-section for the
first region (no acceptance cut, and with the $W^\pm$ leptonic decay branching
ratio included) is :
\begin{displaymath}
  \sigma(pp\to\pstar W^+W^-\pstar\to\pstar\mu^\pm e^\mp\nu\bar\nu\pstar) = 4.0\pm 0.7~\mathrm{fb}.
\end{displaymath}

\begin{figure}[h!]
  \centering
  \begin{subfigure}[b]{.495\textwidth}
    \centering
    \includegraphics[width=\textwidth]{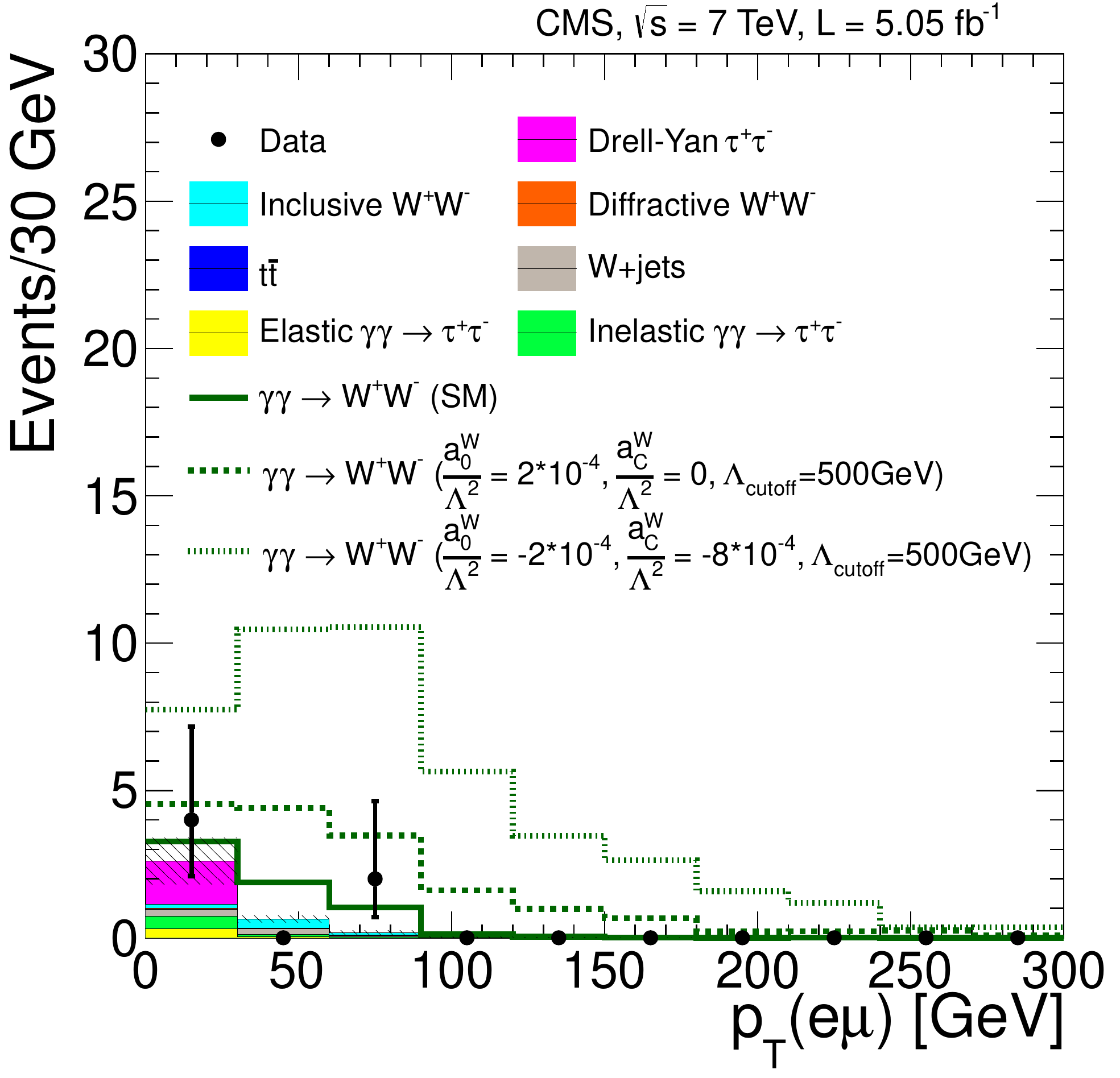}
    \caption{}
    \label{fig:ww-distrib}
  \end{subfigure}
  \begin{subfigure}[b]{.495\textwidth}
    \centering
    \includegraphics[width=\textwidth]{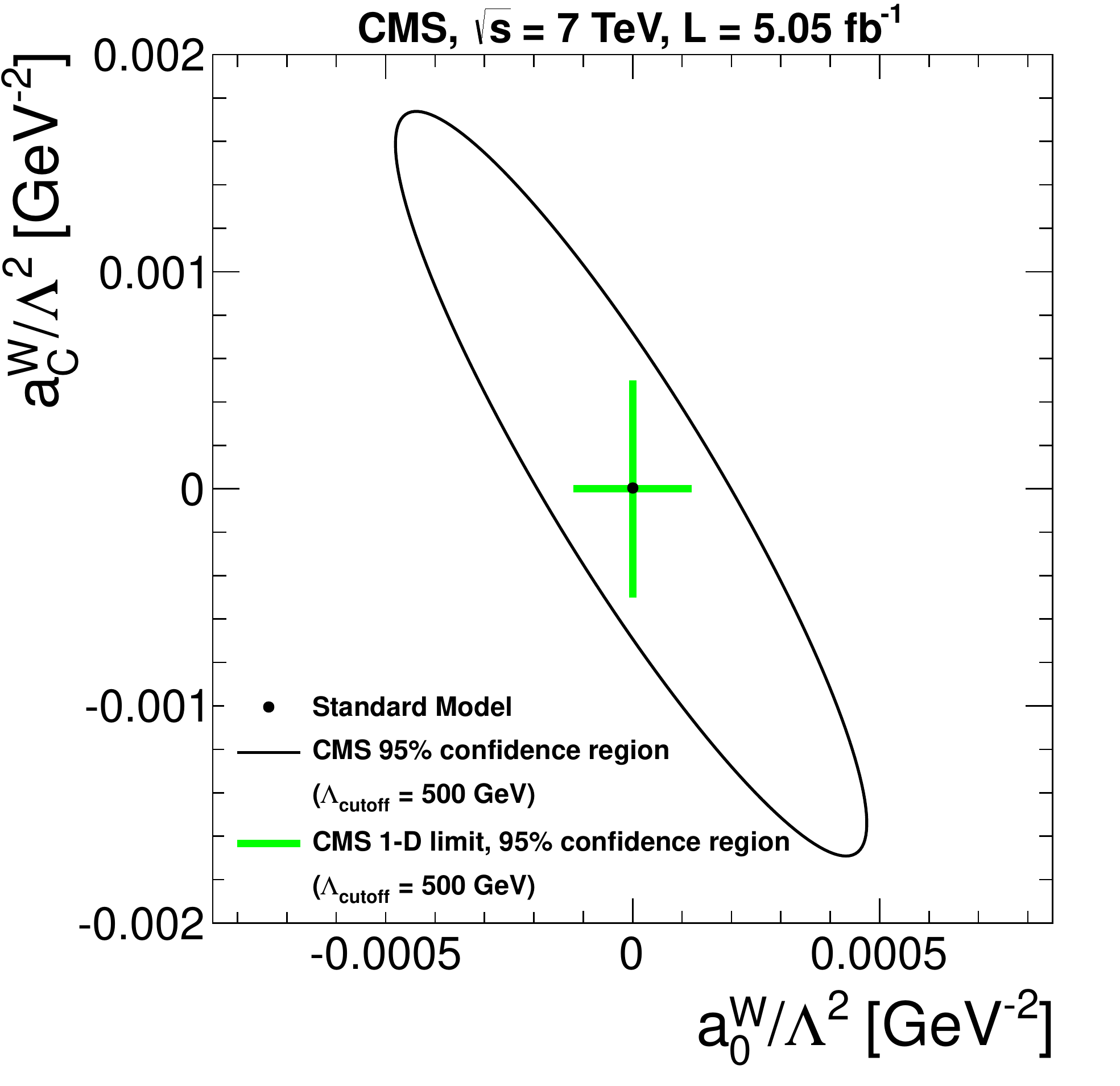}
    \caption{}
    \label{fig:ww-ellipsis}
  \end{subfigure}
  \caption{\small (a) Events passing the full $\gamma\gamma\to W^+W^-$
    selection, with the leptons pairs' transverse momentum relaxed.
    The filled histograms represent the backgrounds while the solid line
    represents the Standard model prediction of such exclusive two-photon
    production of $W^\pm$ pairs, and the dashed ones are two anomalous quartic
    gauge couplings examples given as a mean for comparison.
    (b) One- and two-dimensional limits on the two anomalous quartic gauge
    couplings parameters according to the CMS upper limits on the
    $\gamma\gamma\to W^+W^-$ production cross section.}
\end{figure}

\begin{figure}[h!]
  \centering
  \includegraphics[width=.46\textwidth]{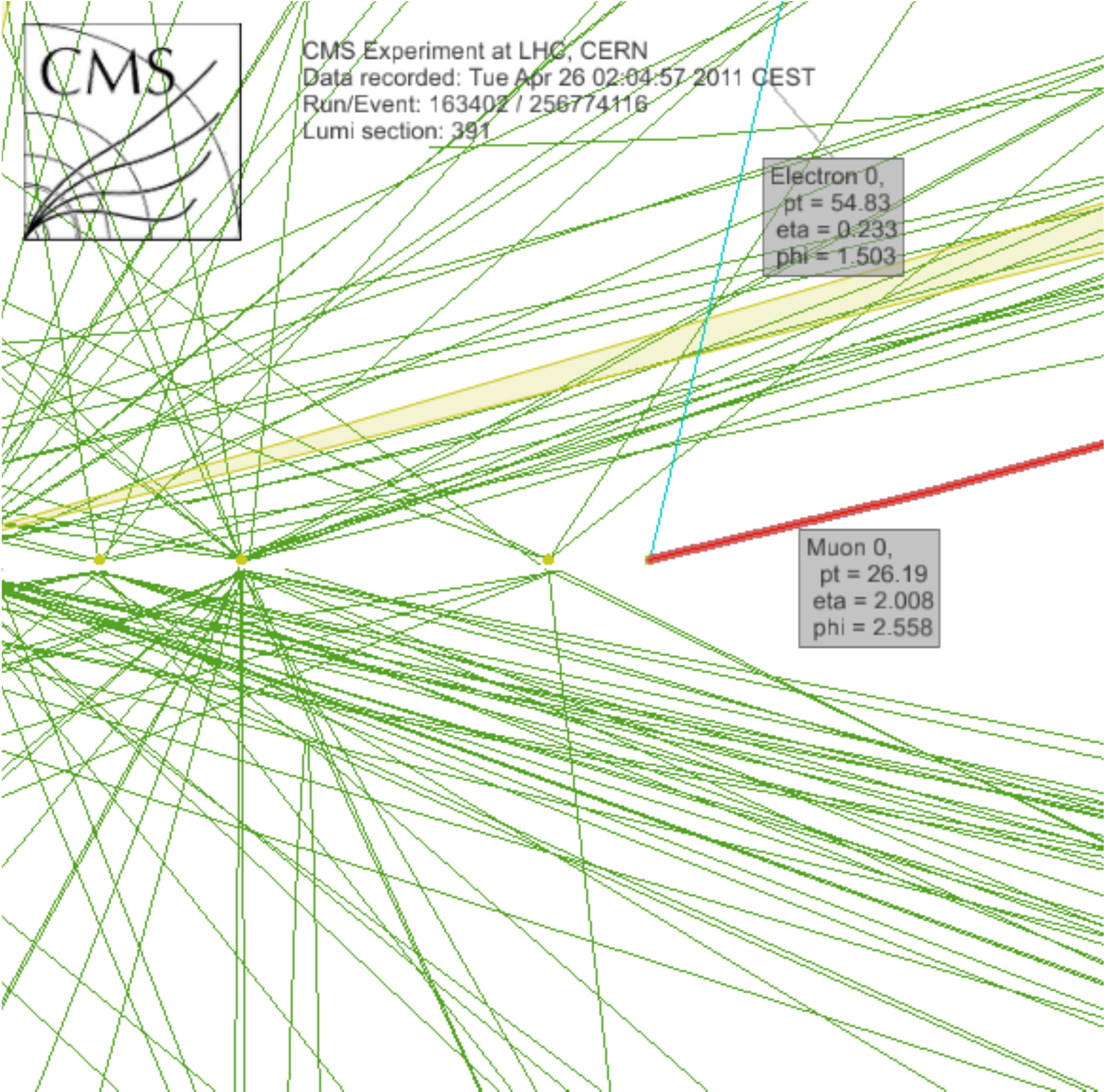}\hspace{.8em}
  \includegraphics[width=.46\textwidth]{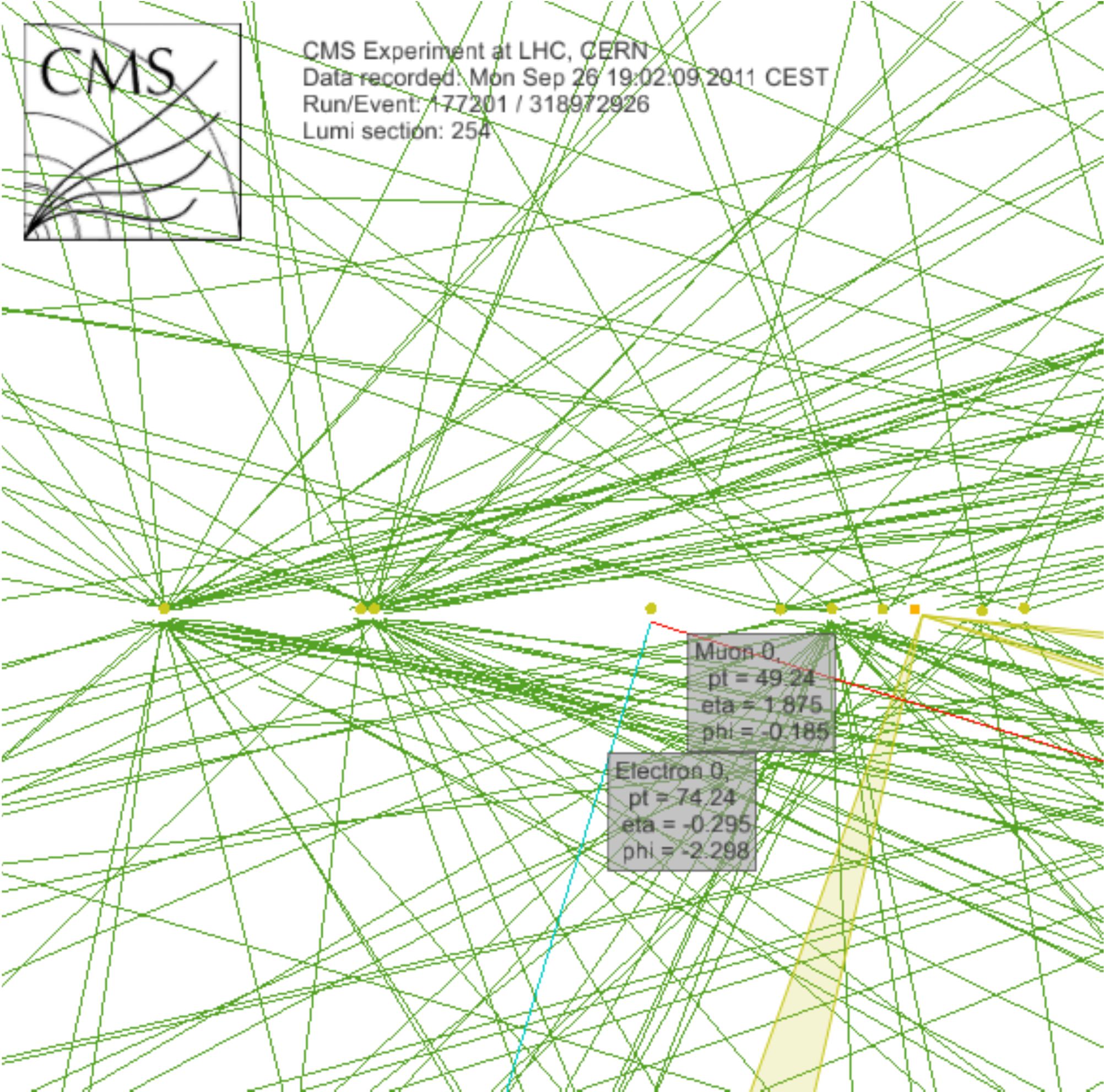}
  \caption{\small Event displays of the two selected Standard model candidates
    are shown in a zoom over the inner tracking system of CMS. The numerous
    vertices present on these figures denote the hard events pileup conditions
    encountered during the 2011 data-taking period.}
  \label{fig:ww-events-display}
\end{figure}

As seen in Fig. \ref{fig:ww-distrib} which depicts the full selected events with
the pair transverse momentum cut relaxed, a total of two
$\gamma\gamma\to W^+W^-$ events candidates (displayed in Fig.
\ref{fig:ww-events-display}) are observed in this Standard model region.

No events are observed in the AQGC search region. An upper limit on the
production cross-section is set according to the theoretical predictions. A
$95\%$ confidence level interval is given for the Poisson mean for signal events
in this window :
\begin{displaymath}
  \sigma\left(p_T(\ell)>20~\mathrm{GeV}, |\eta(\ell)|<2.4, m(e^\pm\mu^\mp)>20~\mathrm{GeV}, p_T(e^\pm\mu^\mp)>100~\mathrm{GeV}\right) < 1.9~\mathrm{fb}.
\end{displaymath}

The two parameters controlling these AQGCs can finally be bounded in a tighter
way than the previous limits by OPAL\cite{Abbiendi:2004bf} and D\O{}
\cite{Abazov:2013opa}. These 1-and 2-dimensional CMS limits at 95\% C.L. are
drawn in Fig. \ref{fig:ww-ellipsis}, and are approximately two orders of
magnitude more stringent than the best limits obtained at the Tevatron, and at
LEP. The one-dimensional bounds are :

\begin{eqnarray*}
|a^{W}_{0}/\Lambda^2| < 1.5\times 10^{-4}~\mathrm{GeV}^{-2},\mathrm{~and~}|a^{W}_{c}/\Lambda^2| < 5.0\times 10^{-4}~\mathrm{GeV}^{-2},~~(\Lambda_\mathrm{cutoff}=500~\mathrm{GeV})\\
|a^{W}_{0}/\Lambda^2| < 4.0\times 10^{-6}~\mathrm{GeV}^{-2},\mathrm{~and~}|a^{W}_{c}/\Lambda^2| < 1.5\times 10^{-5}~\mathrm{GeV}^{-2},~~(\mathrm{no~form~factor})\hspace{1.66em}
\end{eqnarray*}

These results either include or not a dipole form factor with a cutoff scale
$\Lambda_\mathrm{cutoff}=500~\mathrm{GeV}$ to avoid the unitarity violation of
the anomalous models at high two-photon energies. The case with no form factors
is to be taken with care, as they are driven by high-energy two-photon
interactions beyond the unitarity bound.

With two candidates on an undetected channel, the best limits on these anomalous
quartic gauge couplings can be extracted. The limits exceed by two orders of
magnitude the previous results, and are competitive with the current CMS
analyses of such couplings based on tri-boson production \cite{CMS-PAS-SMP-13-009}.

\section{Summary and outlook
\label{sec5}}

In this note, several achievements were shown in the experimental search for
exclusive processes at the LHC.
First, the theoretical predictions for the central-exclusive $\gamma\gamma$
production, mediated by gluons only, are in good agreement with the observed
upper limit on the cross-section.
This result goes hand in hand with the search for the two-photon production of
leptons pairs, which enables to improve the understanding of this purely
electromagnetic process with an observation of 17 candidates for the dielectron
channel and a measurement of a production cross-section at $\sqrt s=7$ TeV
consistent with the theoretical prediction for the dimuon channel.
Finally, with two candidates on a previously undetected $\gamma\gamma\to W^+W^-$
process, the best limits on the anomalous quartic gauge couplings can be
extracted. These limits, while exceeding two orders of magnitude more stringent
results with respect to the previous attempts, are competitive with the current
CMS analyses on such couplings.

The results presented in this note provide evidence both for the excellent
performance of the CMS experiment, and its potential for future measurements of
exclusive processes at the LHC.





\bibliographystyle{ieeetr}
\bibliography{lowx_forthomme}


\end{document}